\documentclass[prb,twocolumn,preprintnumbers,floatfix]{revtex4-1}
\usepackage{graphicx}% Include figure files
\usepackage{dcolumn}% Align table columns on decimal point
\usepackage{bm}% bold math
\usepackage{latexsym,epsfig}
\usepackage{graphicx}
\usepackage{verbatim}
\usepackage{comment}
\usepackage{amsmath}
\usepackage{amssymb}
\usepackage{stmaryrd}
\usepackage{color}
\usepackage{epstopdf}
\usepackage{grffile}
\usepackage{physics}

\DeclareGraphicsExtensions{.eps}
\newcommand{\beq}{\begin{equation}}
\newcommand{\eeq}{\end{equation}}
\newcommand{\bea}{\begin{eqnarray}}
\newcommand{\eea}{\end{eqnarray}}
\newcommand{\ben}{\begin{eqnarray*}}
\newcommand{\een}{\end{eqnarray*}}
\newcommand{\bfig}{\begin{figure}}
\newcommand{\efig}{\end{figure}}

\usepackage{hyperref}
\hypersetup{
    colorlinks=true,      
    urlcolor=blue,
    citecolor=blue,
    linkcolor=blue
}

\newcommand{\tapan}[1]{\textcolor{blue}{\bf #1}}

\definecolor{suman}{rgb}{1,0,1}

\begin{document}
\title{Realizing symmetry protected topological phase through dimerized interactions}

\author{Suman Mondal$^{1,2}$, Ashirbad Padhan$^{1}$ and Tapan Mishra$^{1,3,4}$}
\email{mishratapan@gmail.com}
\affiliation{$^{1}$Department of Physics, Indian Institute of Technology, Guwahati, Assam - 781039, India}
\affiliation{$^{2}$ Institute for Theoretical Physics, Georg-August-Universität Göttingen, Friedrich-Hund-Platz 1, 37077 Göttingen, Germany}
\affiliation{$^3$School of Physical Sciences, National Institute of Science Education and Research, Jatni 752050, India}
\affiliation{$^4$Homi Bhabha National Institute, Training School Complex,
Anushaktinagar, Mumbai 400094, India}

\date{\today}

\begin{abstract}
We show that a one dimensional lattice of spinless fermions can undergo a topological phase transition only due to the effect of interaction. By allowing a dimerized interaction among the particles we show that the system exhibits a symmetry protected topological phase which does not exists in the absence of interaction and the system is a gapless state. The non-trivial topological character appears due to the onset of two degenerate bond-order phases as a function of dimerized interaction which are found to be topologically distinct from each other. As a result a topological phase transition occurs between these bond order phases through a gap closing point. However, in the limit of strong interaction, the topologically distinct bond order phases are separated from each other through a gapped charge density wave phase possessing a local antiferromagnetic order. 
% Interestingly   a trivial to topological phase transition as a function of interaction. In the limit of weak interactions the transition occurs through a gap closing point. The topological phase transition is robust as long as the system does not spontaneously break the translation symmetry due to interaction.or strong enough interactions, the system breaks the translational symmetry and we see a transition between the bond order phase and charge density wave phase, which belongs to the Ising universality class. 
We characterize this emergent topological nature by the edge states, Berry phase and a non-local string order parameter and provide possible experimental signatures in terms of Thouless charge pumping and density-density correlation.

%  At large interactions, by exhibiting the charge density wave phase,  We characterize the nontrivial phase by calculating the Barry phase associated with it using twisted boundary conditions and by looking at the localized edge states. Also, a robust topological charge pumping is shown by periodically varying the system parameters in the topology protected regime.
\end{abstract}

%\pacs{75.40.Gb, 67.85.-d, 71.27.+a }

\maketitle
{\em Introduction.-}
The topological phase transitions have been a topic of paramount interest due to their non-trivial physical properties and for not belonging to the conventional Landau theory of phase transition ~\cite{Hassanreview,Klitzing1980,qi_zhang}. 
%Starting from the seminal observation of quantized Hall conductance in quantum Hall states~\cite{Klitzing1980}, enormous progress has been made to understand the topological phases of matter~\cite{qi_zhang} which do not belong to the conventional Landau theory of phase transition. Characterized by non-trivial physical properties and non-local correlations, topological phases are extremely important in the context of fundamental science and technological applications~\cite{quant_memory,Pesin2012,spintronics,Liu15,Alicea2011}. 
An interesting class of topological phases of matter, known as the symmetry protected topological (SPT) phases~\cite{rachel_review,Senthil_rev} which requires the protection of certain underlying symmetry(ies) of the system for their stability.  Characterised by a finite bulk excitation gap and gapless edge/surface modes, the SPT phases are fundamentally known to be robust against small perturbations. Any transition to another gapped phase due to strong perturbation requires the closing of the excitation gap.  Due to their simplest single particle classification, the SPT phases and transitions have been very well understood in the context of topological insulators and superconductors~\cite{Hassanreview, qi_zhang}. One of the simplest models that exhibits an SPT phase is the one dimensional Su-Schrieffer-Heeger (SSH) model ~\cite{SSH,Asboth_rev} which has been simulated in recent experiments using disparate systems~\cite{Atala2013,Takahashi2016pumping,Lohse2016,Mukherjee2017,Lu2014}. 

On the other hand strong interaction is believed to destroy the non-local order of the SPT phases in certain systems due to spontaneous symmetry breaking. However, a well defined topological character can be favoured due to competing interaction and underlying topology in certain many-body systems such as the Haldane phase in one dimension~\cite{Haldane1983,DallaTorre2006,Luca2013,Fazzini2017,fraxanet}. Several recent studies based on the fermionic SSH models have also revealed that the SPT phases are robust against weak to moderate interaction strength~\cite{Sirker2014,Manmana2012,Wang2015,Santos2018, GuoShen2011, Juenemann2017}. Moreover, strong interactions alone may as well favour the SPT phases under proper conditions which has been discussed both in the fermionic as well as bosonic interacting SSH models~\cite{Grusdt2013,Mondal2020,lewenstein1, lewenstein2,lewenstein3,Jakub2021}.   
Recent experimental realizations of the interaction induced SPT phases ~\cite{browyes,bloch_haldane2021} in utracold atomic systems have brought significant motivation to further investigate the role of interaction in stabilizing the SPT phases. 
% 
% many-body systems. In the context of SPT phase Efforts have been made to study the role of interaction on the SPT phases in recent years.  Although interaction is believed to destroy the non-local order of the SPT phases due to spontaneous symmetry breaking, a well defined topological character can be favored due to competing interaction and underlying topology in certain systems. Notable examples are the topological Mott insulators which have been extensively studied recently in different systems. 
% One of the simplest systems that exhibits interaction induced SPT phase transition is the Su-Schrieffer-Heeger (SSH) model where the ground state undergoes a transition from topologically trivial to non-trivial BO phases that are stabilized due to Peierl's instability. Various other mechanisms have been proposed to realize interaction mediated SPT phase transitions in systems of interacting fermions and bosons in recent years revealing novel interplay between topology and interaction. Due to the recent experimental advancement involving ultracold atoms in optical lattices these phases have been observed recently~\cite{browyes,blochrecent}. 

Arguably, it has been understood that the interaction induced SPT phases in one dimensional lattices require a well controlled deformation of the underlying translational symmetry of the associated non-interacting models. However, in this letter we show that an SPT phase can indeed be possible to stabilize in a one dimensional interacting system whose non-interacting limit is a translationally invariant gapless state. 

Considering a one dimensional lattice of spinless fermions or hardcore bosons at half filling we show that an SPT phase emerges only due to the dimerized or bond-alternating nearest neighbour (NN) interactions which break the translational symmetry of the lattice. Interestingly, this topological character does not inherit from the non-interacting limit of the system rather it stems from the emergent gapped bond order (BO) phases due to the dimerized interactions - a phenomenon reminiscent of the half filled SSH model for spinless fermions. We find that depending on the nature of the dimerization, the system exhibits two topologically distinct BO phases resulting in a topological phase transitions (TPT) as a function of the dimerization. In the regime of weak NN interaction, the TPT occurs through a gap closing point at which the dimerization vanishes (i.e. the point of symmetric NN interaction). However, for strong interactions, the transition occurs through a gapped charge density wave (CDW) region. The non-trivial or topological BO phase is found to possess edge states, quantized Berry phase and a non-local string order parameter. Moreover, as the model proposed in this case deals with NN interactions between the particles, this can in principle be simulated using the ultracold dipolar atoms in optical lattices~\cite{dipolar}. Keeping this in mind, we propose possible experimental signatures in the context of Thouless  charge pumping and density-density correlator. 
{\em Model.-}
For our studies we consider a system of one dimensional nearest-neighbour interacting hardcore bosons with particle-hole symmetry described by the model 
\begin{eqnarray}
H=&-& t \sum_{i} (a_{i}^{\dagger}a_{i+1} + \text{H.c.}) \nonumber\\
&+& V_1 \sum_{i~\in~ odd} \left(n_{i}-\frac{1}{2}\right)\left(n_{i+1}-\frac{1}{2}\right) \nonumber\\
&+& V_2\sum_{i~\in~ even} \left(n_{i}-\frac{1}{2}\right)\left(n_{i+1}-\frac{1}{2}\right).
\label{eq:ham}
\end{eqnarray}
Here, $a^\dagger_i(a_i)$ is the creation (annihilation) operator of the particles at site $i$. $V_1$ and $V_2$ are the NN interactions between alternate bonds. We consider dimerized interactions which is achieved by setting $V_1\neq V_2$. When $V_1 = V_2$ ($V_1\neq V_2$), we call the system symmetric (dimerized). Note that the model shown above can be mapped to spin polarized fermions and spins under proper transformations~\cite{mishratvvp}, hence, the observables considered for our studies possess identical behaviour for all the three cases. We perform the simulations using the Density matrix renormalization group (DMRG) method where we consider up to $500$ bond dimensions. Our focus is on the physics at half filling i.e. $\rho=N/L=0.5$, where $N$ and $L$ are the number of particles and system size respectively. We have considered system sizes up to $L = 700$ to avoid finite size effects and set $t=1$ as the energy scale of the system.  

It is well known that at half filling ($\rho=0.5$) the model(\ref{eq:ham}) in the symmetric interaction limit i.e. $V_1=V_2=V$, is the well known $t-V$ model which exhibits a gapless superfluid (SF) to gapped CDW phase transition at the critical point $V = 2t$. In this paper we show that by moving away from this symmetric limit (i.e. for finite dimerization) interesting features appear in the phase diagram leading to the topological phase transition which will be discussed in detail in the following.

%%%%%%%%%%%%%%%%%%%%%%%%%%%%%%%%%%%%%
% FIGURE 1
\begin{figure}[!t]
\begin{center}
\includegraphics[width=1\columnwidth]{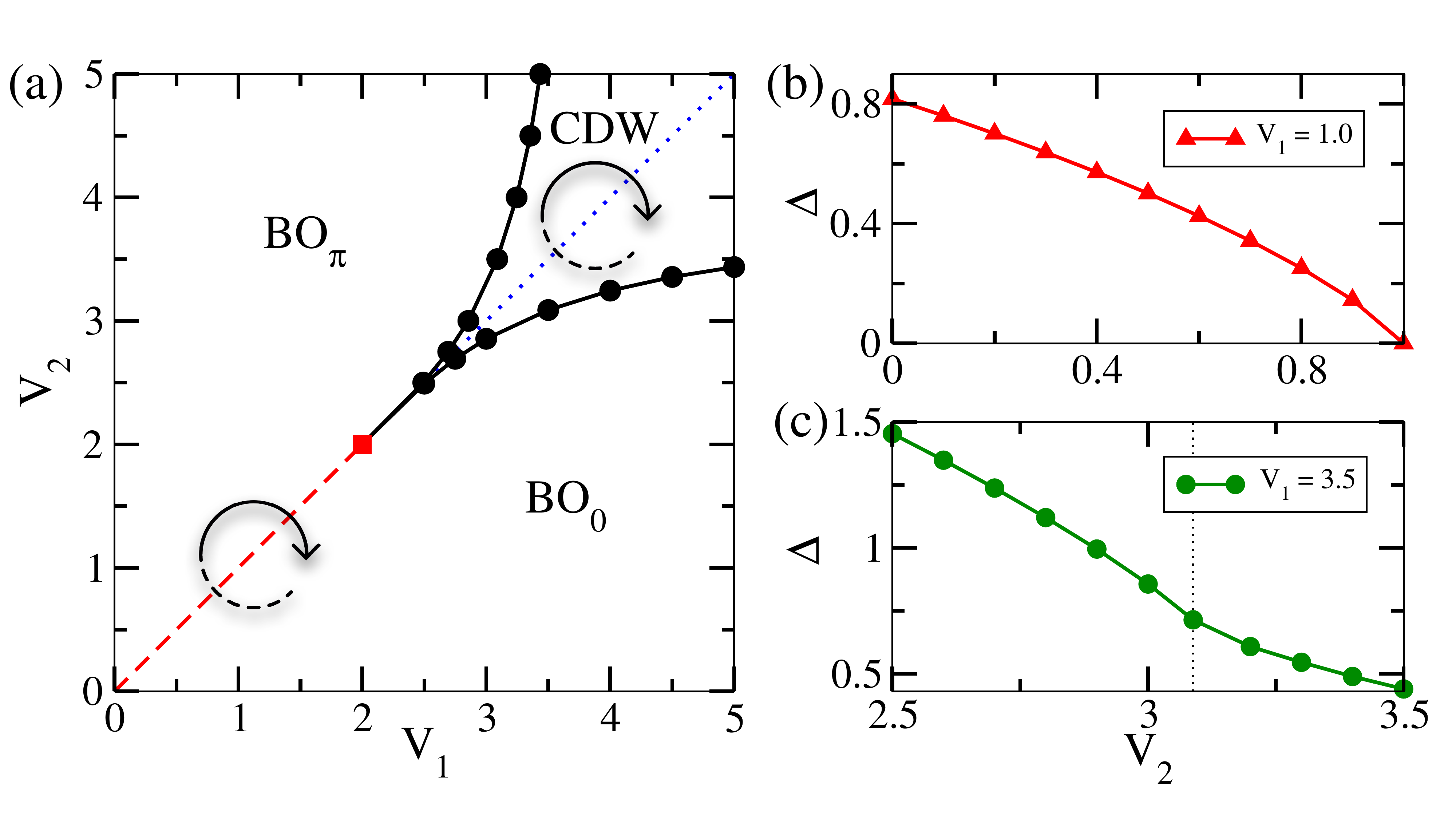}
 \end{center}
% \vspace*{-0.5cm}
\caption{(a) The phase diagram of model~(\ref{eq:ham}) in $V_1-V_2$ plane. The red dashed (blue dotted) line represents the gapless SF (gapped CDW) phase at $V_1 = V_2$. The SF-CDW critical point is marked by the solid red square. The black solid circles represents the boundary between the BO and the CDW phases. The arrow directions in (a) show the pumping protocols. Here $\theta$ in BO$_\theta$ is the associated Berry phase. (b) and (c) show the extrapolated gap at $V_1 = 1.0$ and $3.5$ respectively with varying $V_2$. The thin dotted line in (c) marks the BO-CDW transition point. }
\label{fig:pd}
\end{figure}
%%%%%%%%%%%%%%%%%%%%%%%%%%%%%%%%%%%%%
{\em Phase diagram.-}
Before discussing the topological aspects of our studies, we first present the complete ground state phase diagram of the model~ (\ref{eq:ham}) obtained using our DMRG simulations in Fig.~\ref{fig:pd}(a). As already mentioned before, when $V_1 = V_2$, a transition takes place from the gapless SF phase (red dashed line) to the gapped CDW phase (blue dotted line) at the critical point $V_1 = V_2 = 2$ denoted as the red solid square in Fig.~\ref{fig:pd}(a).  Interestingly, starting from the SF region, as soon as the dimerization is turned on i.e. when $V_1 \neq V_2$, the system immediately becomes gapped. However, the gap which exists at $V_1=V_2 > 2$ remains finite as a function of dimerization. In Fig.~\ref{fig:pd}, we plot the extrapolated values of the charge gap~\cite{mishratvvp} $\Delta_{L\to\infty} = E(L, N+1)+E(L, N-1) -2E(L, N)$ as a function of $V_2$ at two different cuts through the phase diagram i.e. at $V_1=1.0$ (Fig.~\ref{fig:pd}(b)) and $V_1=3.5$ (Fig.~\ref{fig:pd}(c)). The behaviour clearly shows that for $V_1=1$,  the gap closes exactly at $V_2=1$ and remains finite otherwise. On the other hand for $V_1=3.5$, the gap always remains finite as a function of $V_2$. These features indicate that the regions depicted in the entire phase diagram are gapped except the SF phase (red dashed line) at $V_1=V_2$. 

\begin{figure}[t]
\begin{center}
\includegraphics[width=1\columnwidth]{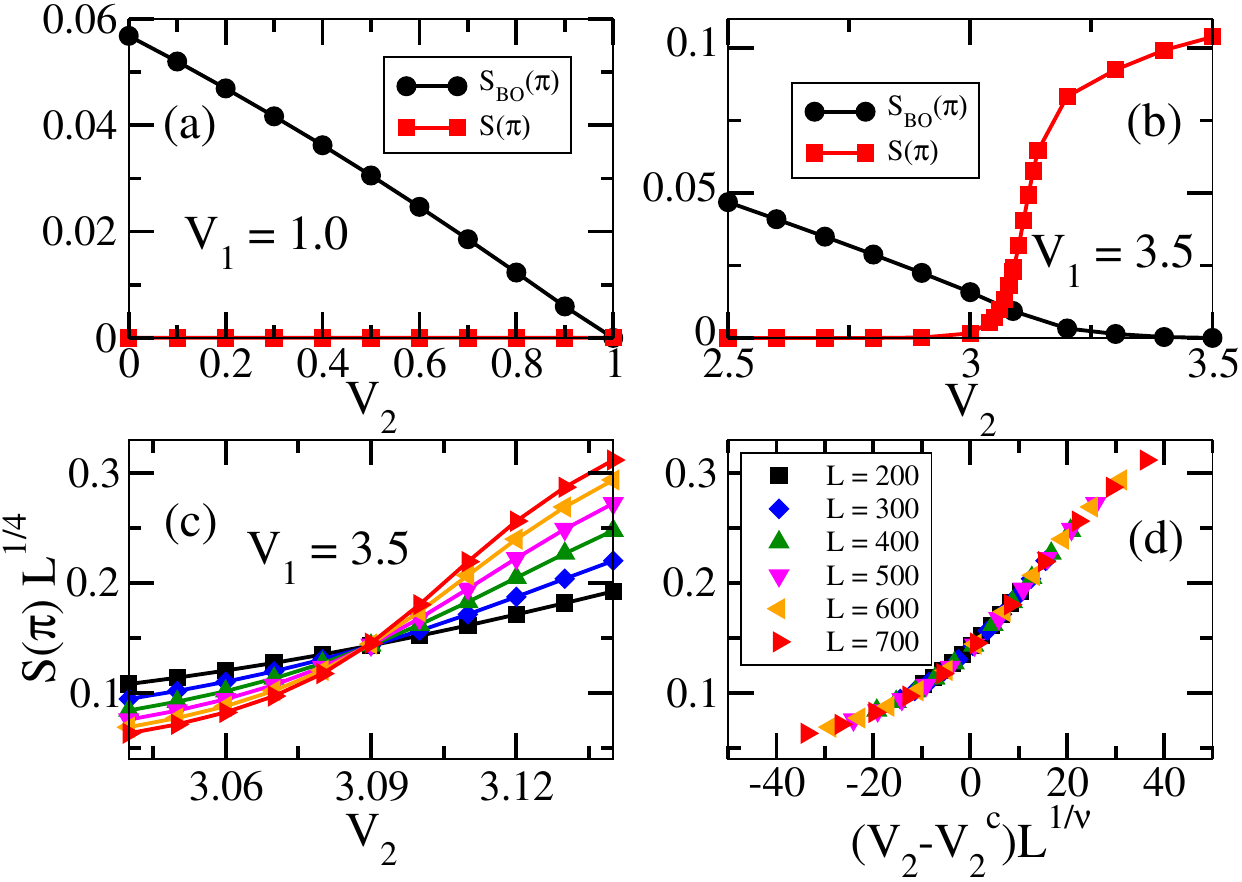}
 \end{center}
% \vspace*{-0.5cm}
\caption{(a) and (b) show the behavior of
$S_{BO}(\pi)$ (black circles) and $S(\pi)$ (red squares) with varying $V_2$ for cuts through $V_1 = 1.0$
and $3.5$ respectively in the phase diagram. (c) The finite-size scaling of $S(\pi)$ is shown
for different values of $L$ at $V_1 = 3.5$ with varying $V_2$. The crossing of the curves for different $L$ at a point represents the BO-CDW transition. (d) Scaled $S(\pi)$ is plotted against the scaled $V_2$ showing the collapse of all points onto a single curve and thus confirming $V_{2}^c=3.09$ as the transition point.}
\label{fig:sfbo}
\end{figure}

Interestingly, the gapped regions immediately below and above the symmetric line (where $V_1\neq V_2$) up to $V_1=V_2=2$ (red square in Fig.~\ref{fig:pd}) are found to possess finite bond ordering induced by the NN interaction exhibiting finite oscillation in the bond kinetic energy $B_i = a_{i}^{\dagger}a_{i+1} + H.c.$ with $i$ denoting the site index. On the other hand, in the limit of strong interactions i.e. $V_1=V_2 > 2$, the CDW nature is found to remain stable up to certain range of dimerization before the system undergoes a transition to the BO phase. While the BO phases are characterized by a finite peak in the bond-order structure factor $S_{BO}(k) = \frac{1}{L^2}\sum_{i,j}e^{ikr}\langle B_{i}B_{j}\rangle$ ~\cite{mishrattpv,mishrattpvvp,mishratvvp,Sansone2009,Ejima2007}, the CDW phase is characterized by a finite peak in the density structure factor $S(k) = \frac{1}{L^2}\sum_{i,j}e^{ikr}(\langle n_{i}n_{j}\rangle - \langle n_{i}\rangle \langle n_{j}\rangle)$. 
%Note that here $\langle n_{i}n_{j}\rangle$ is the density-density correlation function. 
In Fig.~\ref{fig:sfbo} we compare the behaviour of the finite size extrapolated values of $S_{BO}(\pi)$ and $S(\pi)$ by plotting them as a function of $V_2$ at two different cuts through the phase diagram i.e. $V_1=1.0$ and $3.5$. It is clearly seen from Fig.~\ref{fig:sfbo}(a) that for $V_1=1$, the $S_{BO}(\pi)$ remains finite for all the values of $V_2$ except at the symmetric point $V_1=V_2=1$ where $S_{BO}(\pi)$ vanishes.  However, for the cut through $V_1=3.5$, there is a clear indication of transition between the BO and the CDW phases at a critical $V_2$ as shown in Fig.~\ref{fig:sfbo}(b). The BO - CDW phase transition is found to follow the Ising universality class and the critical points are obtained by using appropriate scaling behaviour of the CDW structure factor 
\begin{equation}\label{eq:scale}
S(\pi)L^{2\beta/\nu} = F((V_2-V_2^c) L^{1/\nu})
\end{equation}
where $\beta$ and $\nu$ are the critical exponents~\cite{mishratvvp}. In Fig.~\ref{fig:sfbo}(c) we plot $S(\pi)L^{2\beta/\nu}$ as a function of $V_2$ for $V_1=3.5$ with $\beta = 1/8$ and $\nu = 1$. The curves for different system sizes cross at the critical point which is found to be $V_2^c\sim 3.09$. Now by using the scaling function shown in Eq.~\ref{eq:scale} a perfect data collapse is obtained with $V_2^c=3.09$ as shown in Fig.~\ref{fig:sfbo}(d). Using this approach we trace out the entire CDW phase which is flanked by the BO phases and the BO - CDW transition is marked by the black circles in Fig.~\ref{fig:pd}(a). 
Note that the BO - CDW transition occurs in both the regimes of dimerization i.e. $V_1 < V_2$ and $V_1 > V_2$. 
In the following we will show that the emergence of such bond order phases in such a simple model is responsible for the topological phase transition in the  system.
%%%%%%%%%%%%%%%%%%%%%%%%%%%%%%%%%%%%%
% FIGURE 2
%%%%%%%%%%%%%%%%%%%%%%%%%%%%%%%%%%%%%
% In the following we will show that when $V_1> V_2$, the BO is a trivial phase (denoted by BO$_0$) and when $V_1<V_2$ it is non-trivial/topological (denoted by BO$_\pi$), where $0$ of $\pi$ are the associated \tapan{winding numbers}. The details of the different phases and phase transitions are discussed below in great detail.

%%%%%%%%%%%%%%%%%%%%%%%%%%%%%%%%%%%%%
% FIGURE 3
\begin{figure}[t]
\begin{center}
\includegraphics[width=1\columnwidth]{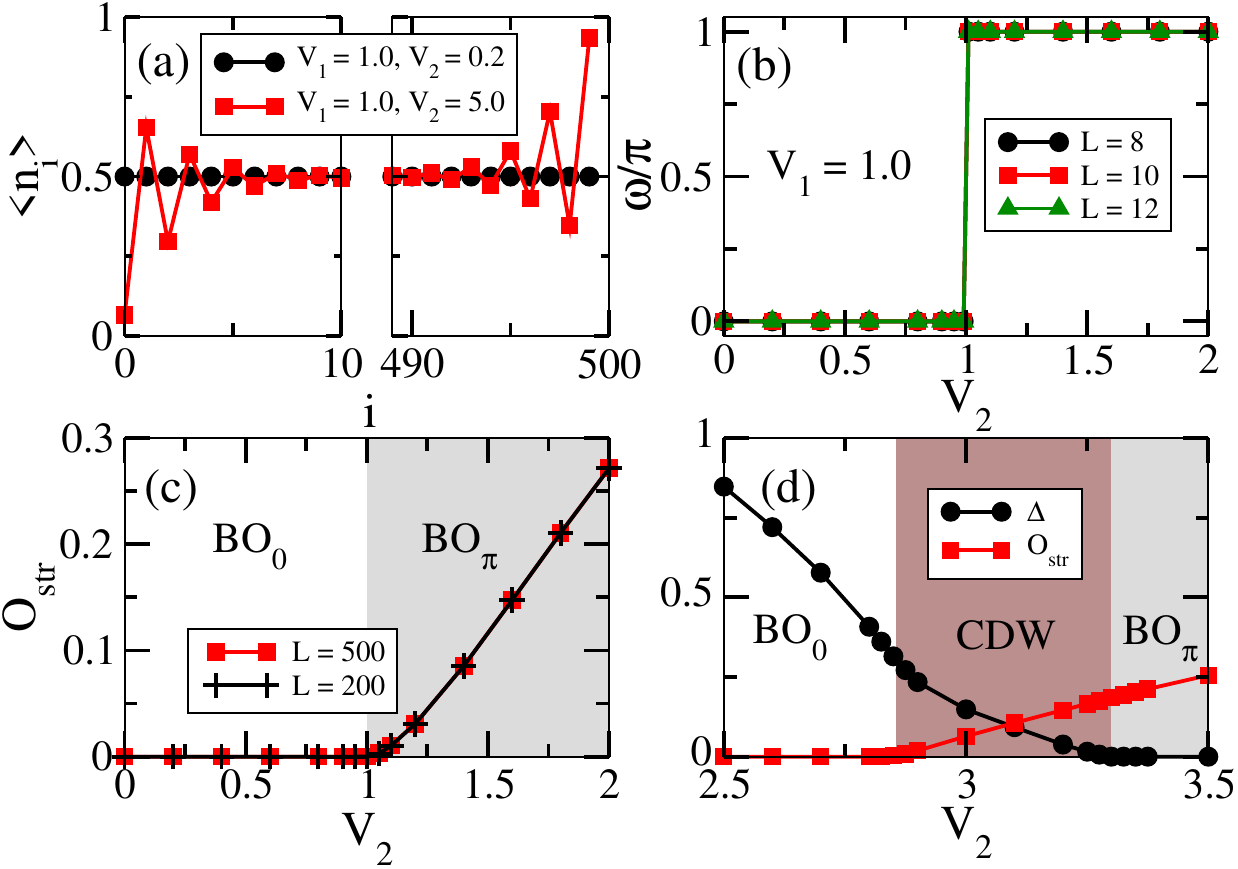}
 \end{center}
% \vspace*{-0.5cm}
\caption{(a) Onsite particle densities $\langle n_i\rangle$ vs $i$ are plotted in the BO$_0$ (black circle) and BO$_\pi$ (red squares) phases. (b) The Berry phase $\omega/\pi$ for different lengths $L=8,~10,~12$ obtained using the ED method is plotted as a function of $V_2$  across the topological phase transition at $V_1 = 1.0$. (c) $O_{str}$ vs $V_2$ for $V_1 = 1.0$ are plotted for $L=500$ (red squares) and $L=200$ (black plus) where the effective topological phase transition is valid. (d) The extrapolated values of $\Delta$ (black circles) and $O_{str}$ for $L=400$ are plotted as a function of $V_2$ for $V_1=3.0$. The finite (vanishing) values of $O_{str}$ ($\Delta$) after $V_2=3.3$ indicate a transition to the BO$_\pi$ phase.}
\label{fig:edge}
\end{figure}
%%%%%%%%%%%%%%%%%%%%%%%%%%%%%%%%%%%%%

{\em Topological Character.-}
After discussing the bulk phase diagram of the model, in this part we discuss the topological character of these BO phases and the transition between them. As already highlighted in the bulk phase diagram, there exist two BO phases across the symmetric line of interaction. Here we show that these BO phases are topologically different from each other and a phase transition occurs between them through the gap closing point at $V_1=V_2$  along the red dashed line that extends up to $V_1=V_2=2$ in the phase diagram of Fig.~\ref{fig:pd}. 

As a first signature to distinguish between the two BO phases, we examine the edge properties of a finite system in search of the degenerate zero energy edge states which are the typical character of the topological insulators. To this end we plot the real space onsite particle numbers at each site for $V_1 > V_2$ (black circles) and $V_1 < V_2$ (red squares) in Fig.~\ref{fig:edge}(a). The polarized edge population for $V_1 < V_2$ in the figure indicates the topological character of the BO phase. To further quantify this feature we compute the Berry phase ($\omega$) as the topological invariant in such interacting system. By using the twisted boundary condition~\cite{Hatsugai} i.e. by setting $a_{i} \to {\rm e}^{i\phi / L} a_i$ and by varying the twist angle $\phi$ from $0$ to $2\pi$ the Berry phase can be computed as;
\begin{equation}
\omega = \int_0^{2\pi} d\phi \langle \psi(\phi) |\partial_\phi \psi(\phi) \rangle.
\end{equation}
We calculate $\omega$ for small systems of size up to $L=12$ sites using the exact diagonalization (ED) method and plot them with respect to $V_2$ for $V_1 = 1.0$ in Fig.~\ref{fig:edge}(b). The discontinuous jumps from $\omega=0$ to $\omega=\pi$ at $V_1=1$ for all lengths indicate a clear signature of topological phase transition. With this distinction, we classify the BO phases below and above the symmetric line as trivial and topological BO phases denoted by BO$_0$ and BO$_{\pi}$ phases respectively as depicted in the phase diagram of Fig.~\ref{fig:pd}(a). 
%Similar features can also be seen across the entire SF line (see Fig.~\ref{fig:edge}(c)). 

%%%%%%%%%%%%%%%%%%%%%%%%%%%%%%%%%%%%%
% % FIGURE 4
% \begin{figure}[t]
% \begin{center}
% \includegraphics[width=1\columnwidth]{string_gap.pdf}
%  \end{center}
% % \vspace*{-0.5cm}
% \caption{ The finite-size scaling of $S(\pi)$ is shown for different values of $L$ at $V_1 = 3.5$ with varying $V_2$. The clean crossing of the curves for different $L$ at a point represents the transition point for BO to CDW phase with the critical exponent $\beta = 1/8$ (see the text)}
% \label{fig:scale}
% \end{figure}
%%%%%%%%%%%%%%%%%%%%%%%%%%%%%%%%%%%%%

Furthermore, we show that the topological BO$_\pi$ phase is found to exhibit a non-local string order which can be quantified by the string order parameter~\cite{den_nijsPRB1989, tasakiPRL1991, hadaPRB1992} defined as
\begin{equation}\label{eq:ostr}
O_{str}(r)=  -\langle z_i  e^{i \frac{\pi}{2} \sum_{k=i+1}^{j-1} z_k}  z_{j}\rangle, 
\end{equation}
where $z= 1- 2a^\dagger a$ and $r=|i-j|$. To avoid the edge sites and to consider a maximum distance $r$, we chose $i = 2$ and $j=L-1$ in Eq.~\ref{eq:ostr}.
%\begin{equation}
%O_{str}=  -\langle z_2  e^{i \frac{\pi}{2} \sum_{j=3}^{L-2} z_{j}}  z_{L-1}\rangle.
%\end{equation}
We plot $O_{str}$ as a function of $V_2$ for $V_1 = 1.0$ in Fig.~\ref{fig:edge}(c) for two different lengths such as $L=200$ (black plus) and $500$ (red squares) which becomes finite after the critical point of transition i.e. at $V_1 = V_2$. This feature clearly indicates the existence of a non-local string order in the BO$_{\pi}$ phase and is an important signature which distinguishes the two BO phases having the same bulk properties in the regimes of $V_1 > V_2$ and 
$V_1 < V_2$. However, across the transition through the CDW phase, due to construction, the $O_{str}$ is expected to be finite in the CDW phase as well which can be seen in Fig.~\ref{fig:edge}(d) where the $O_{str}$ (red squares) is plotted as a function of $V_2$ for $V_1=3.0$.  
%In Fig.~\ref{fig:edge}(d), we plot $O_{str}$ as a function of $V_2$ at $V_1 = 3.0$ where $O_{str}$ becomes finite as soon as the system enters into the CDW phase. 
Therefore, in order to distinguish the BO phases from the CDW phase in this case we plot the excitation gap $\Delta$ (black circles) as a function of $V_2$ which vanishes in the BO$_\pi$ phase signifying the existence of gapless edge states. Comparing the values of $O_{str}$ and $\Delta$ one can clearly separate the CDW phase (brown shaded region) from the two BO phases and the transition between them.

%%%%%%%%%%%%%%%%%%%%%%%%%%%%%%%%%%%%%
% FIGURE 6
% \begin{figure}
% \begin{center}
% \includegraphics[width=1\columnwidth]{density_corrs.pdf}
%  \end{center}
% % \vspace*{-0.5cm}
% \caption{(color online) (a) String order parameter ($C$) is plotted for $V_1=1$ with varying $V_2$ where the effective topological phase transition is valid. We see that when $V_1>V_2$ the $C$ becomes finite, depicting the topological phase transition. (b) The $C$ is plotted for $V_1 = 3.5$ ($V_2 = 3.5$) with varying $V_2$ ($V_1$) denoted by solid black line with circles (solid red line with squares). The thin dotted line marks the phase transition point for BO$_\theta$ and CDW phase. We can see that the $C$ is finite for both the BO$_\pi$ and CDW phase. Note that in this case, the TPT is not valid because of the symmetry breaking.}
% \label{fig:str}
% \end{figure}
%%%%%%%%%%%%%%%%%%%%%%%%%%%%%%%%%%%%%

From the above discussion, it is evident that the presence of only dimerized interaction is sufficient to establish a topological BO phase and a topological phase transition. This topological phase transition is protected by the emergent bond inversion symmetry and the already assumed particle-hole symmetry of the system. An interesting inference which can be drawn from this analysis is that a topological transition is not possible between the two BO phases beyond the limit $V_1 = V_2 = 2$. This ambiguity can be attributed to the spontaneously broken translational as well as the bond inversion symmetry in the gapped CDW phase. Moreover, due to the  particle-hole symmetry, the CDW phase also exhibits non-degenerate and energetic edge states (not shown) as mid gap states.
% 
% \tapan{On the other hand it is not possible to adiabatically connect the trivial and nontrivial BO phases beyond the $V_1 = V_2 >2$. This ambiguity can be attributed to the spontaneously broken translational symmetry by the CDW phase which is gapped. }
% So the bulk-boundary correspondence can only be valid for $V_1$ or $V_2<2$ and our discussion about the topological phases will be restricted to that regime.  
%%%%%%%%%%%%%%%%%%%%%%%%%%%%%%%%%%%%%
% FIGURE 5
\begin{figure}[!t]
\begin{center}
\includegraphics[width=1\columnwidth]{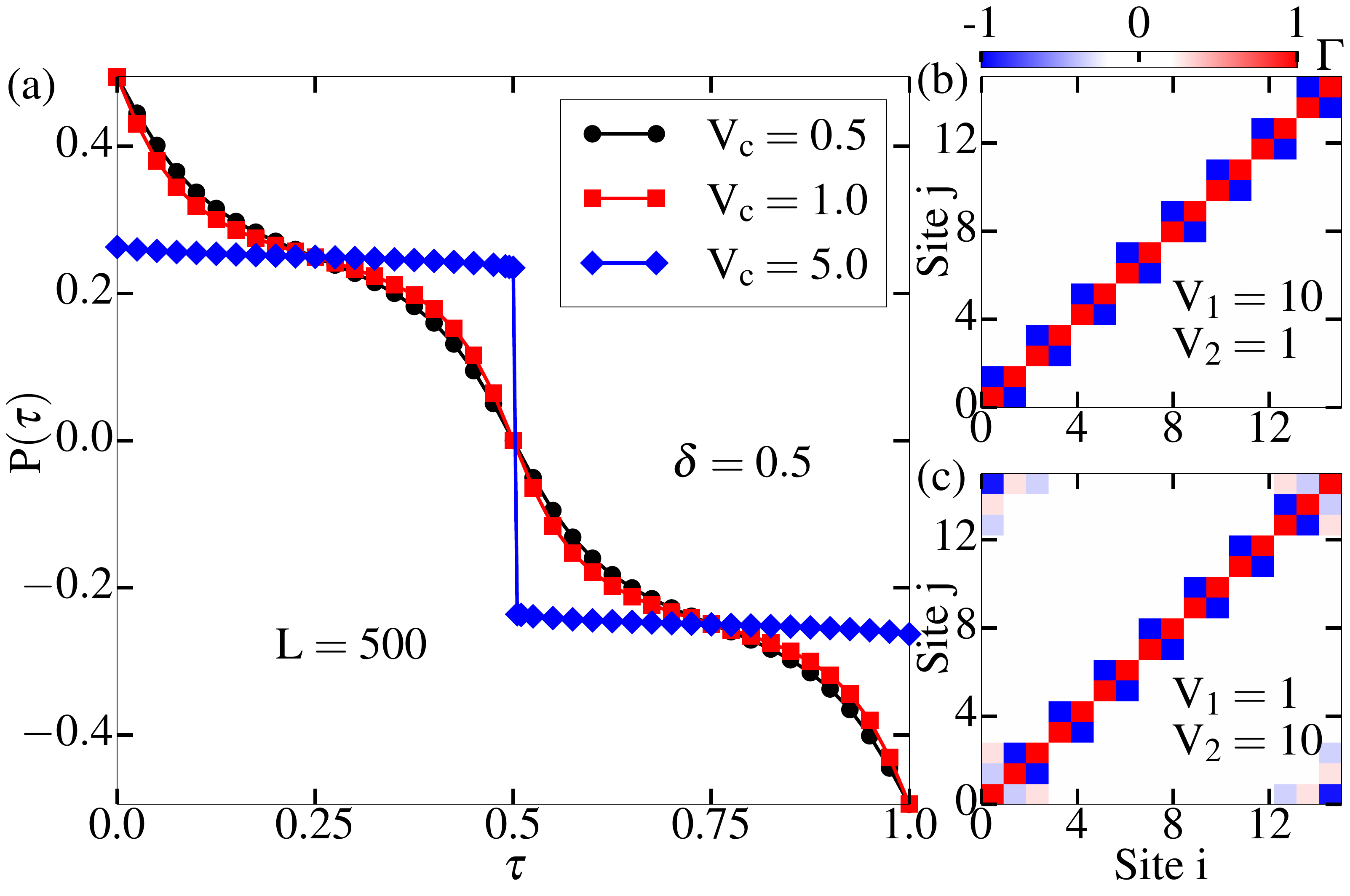}
 \end{center}
% \vspace*{-0.5cm}
\caption{(a) Shows the evolution of polarization following a pumping cycle across three different critical 
points such as $V_c = 0.5$ (black circles), $1.0$ (red squares) and  $V_c=5.0$ (blue diamonds) of the phase diagram in Fig.~\ref{fig:pd} obtained using the DMRG simulations. The density correlation matrix $\Gamma$ obtained using the ED method is plotted for $L=16$ sites in (b) for the BO$_0$ phase ($V_1=10.0, V_2=1.0$) and in (c) for the BO$_{\pi}$ phase ($V_1=1.0, V_2=10.0$).}

\label{fig:rm}
\end{figure}
%%%%%%%%%%%%%%%%%%%%%%%%%%%%%%%%%%%%%

{\em Experimental realization.-}
In this part, we provide the possible signatures of the topological character of the BO$_\pi$ phase in the context of the Thouless charge pumping (TCP)~\cite{chargepumping} which has been successfully observed in different experiments to characterise the topological nature of non-interacting systems~\cite{Takahashi2016pumping,Lohse2016}. The recent generalization of the TCP to interacting systems are based on systems with finite hopping dimerization and are described by the Rice-Mele models~\cite{Kuno2021,Kuno2020,Hayward2018,Mondal2020}. In general the pumping protocol relies on the adiabatic variation of a parameter which is responsible for the TPT. Moreover, an additional symmetry breaking term that helps to keep the gap of the system open in the pumping cycle is a necessary condition. In the process the number of charge quanta pumped in a cycle reveals a topological invariant.

%which is an interesting transport phenomenon exhibited by the topological systems. By periodically varying the system parameter, a quantized number of charges can be pumped in each pumping cycle. This phenomenon is prevalent and

Here, we propose a pumping protocol based on the Hamiltonian (Eq.~\ref{eq:ham}) with an  additional symmetry breaking parameter $\Delta$, given by
\begin{align}
{H}_{\rm p} =&- t \sum_{i} (a_{i}^{\dagger}a_{i+1} + \text{H.c.}) \nonumber\\
& -\sum_i (V_c - (-1)^i \delta  \cos(2\pi \tau))\left(n_{i}-\frac{1}{2}\right)\left(n_{i+1}-\frac{1}{2}\right)\nonumber\\
&+ \Delta \sin(2\pi \tau) \sum_i (-1)^i n_i\,,
\label{eq:pump}
\end{align}
where $\tau$ is the pumping parameter. The parameter $\Delta$ ensures that the gap remains open during the pumping cycle, especially at $\tau = 1/4$ and $3/4$. Note that in the limit $\Delta = 0$, the above Hamiltonian reduces to model~\ref{eq:ham} with $V_1 < V_2$ ($V_1 
> V_2$) at $\tau = 0$ ($\tau = 1/2$). The $V_c$ is the gap closing point that resides on the gapless line in the phase diagram of Fig.~\ref{fig:pd} when $\delta = \Delta = 0$ (i.e., $V_1=V_2=V_c$). %Since the effective topological phase transition is valid before the CDW order appears, we consider  $V_c$ less than $ 2$. 

% %%%%%%%%%%%%%%%%%%%%%%%%%%%%%%%%%%%%%
% % FIGURE 7
% \begin{figure}
% \begin{center}
% \includegraphics[width=1\columnwidth]{RM.pdf}
%  \end{center}
% % \vspace*{-0.5cm}
% \caption{(color online) The figure shows the evolution of polarization in a pumping cycle. Here we consider two separate points of  $V_c = 0.5$ (solid line with black circles) and $1.0$ (solid line with red squares), $\delta = \Delta = 0.5$ and a finite system of length $L=500$. }
% \label{fig:rm}
% \end{figure}
% %%%%%%%%%%%%%%%%%%%%%%%%%%%%%%%%%%%%%

To study the TCP, we consider different values of $V_c$ with $\delta = \Delta = 0.5$ and compute the polarization given by
\begin{equation}
P(\tau) = \frac{1}{L} \sum_{i=0}^{L-1} \langle \psi(\tau) | (i-i_0) n_i | \psi(\tau) \rangle
\end{equation}
where $i_0 = (L-1)/2$.
The total number of charge pumped during the pumping cycle is then given as $Q = \int_0^1\partial_{\tau}P(\tau)d\tau$.
Here, $\ket{\psi(\tau)}$ is the ground state  wavefunction for a particular value of $\tau$. We compute $P(\tau)$ for a system of size $L = 500$ and plot them as a function of $\tau$ in Fig.~\ref{fig:rm}(a) for $V_c=0.5$ (black circles) and $1$ (red squares). The continuous change of $P(\tau)$ from $0.5$ to $-0.5$ indicates a robust charge pumping of one particle \tapan{$Q=1$} in a pumping cycle. However, for a critical $V_c=5.0$ (blue diamonds) a clear breakdown of charge pumping is seen indicating no SPT transition. Furthermore, we also examine the density-density correlation matrix $\Gamma=\langle z_iz_j\rangle-\langle z_i\rangle\langle z_j\rangle$, which can be accessed in the quantum gas experiments~\cite{browyes} to distinguish between the trivial and the topological phases. The two isolated red points at the two opposite ends of the correlation matrix $\Gamma$ in Fig.~\ref{fig:rm}(c) clearly confirm the presence of the edge states in the BO$_\pi$ phase that distinguishes it from the BO$_0$ phase (see Fig.~\ref{fig:rm}(b)) where the edge states are absent.

{\em Conclusions.-} 
We have shown that for a one dimensional model of spinless fermions or hardcore bosons, a symmetry protected topological phase can be achieved by allowing only dimerized NN interactions while the hopping remains uniform. By extensive DMRG analysis we show that a trivial to topological phase transition occurs as a function of dimerized interaction which stems from the emergent BO phases in the system. While there exists an SPT phase transition in the limit of weak NN interaction, for strong interaction, the two BO phases are separated by a gapped CDW phase indicating no SPT transition. The model considered in our analysis is one of the simplest systems to exhibit interaction induced topological phase transition and can be simulated using the ultracold dipolar atoms in optical lattice. Keeping this in mind, we provide possible experimental signatures of the  topological phases in terms of Thouless charge pumping and density-density correlation function. 

Our analysis provides a route to obtain interaction induced SPT phase transition for one dimensional spinless fermions or hardcore bosons without explicitly constraining the particle tunneling and thereby breaking the translational symmetry. This finding therefore provides an interesting platform to study the robustness of the topological character in presence of other perturbations. For example, an immediate extension can be to see the effect of diagonal disorder and frustrated next-nearest neighbour hoppings on the SPT phases. Furthermore, it would be interesting to explore the stability of the topological character by moving away from the one dimensional limit in coupled one dimensional lattices.

\bibliography{references}

\end{document}